\documentclass[12pt,prd,aps,amssymb,amsmath,tightenlines,showpacs]{revtex4}
\begin{document}

\def\half{\textstyle{\frac{1}{2}}}
\def\cP{\mathcal P}
\def\cC{\mathcal C}
\def\cT{\mathcal T}

\topmargin=0.0cm

\title{Quantum counterpart of spontaneously broken classical $\cP\cT$ symmetry}

\author{Carl~M.~Bender${}^1$}
\email{cmb@wustl.edu}

\author{Hugh~F.~Jones${}^2$}\email{h.f.jones@imperial.ac.uk}

\affiliation{${}^1$Department of Physics, Washington University, St. Louis
MO 63130, USA}

\affiliation{${}^2$Blackett Laboratory, Imperial College, London SW7 ABZ, UK}

\begin{abstract}
The classical trajectories of a particle governed by the $\cP\cT$-symmetric
Hamiltonian $H=p^2+x^2(ix)^\epsilon$ ($\epsilon\geq0$) have been studied in
depth. It is known that almost all trajectories that begin at a classical
turning point oscillate periodically between this turning point and the
corresponding $\cP\cT$-symmetric turning point. It is also known that there are
regions in $\epsilon$ for which the periods of these orbits vary rapidly as
functions of $\epsilon$ and that in these regions there are isolated values of
$\epsilon$ for which the classical trajectories exhibit spontaneously broken
$\cP\cT$ symmetry. The current paper examines the corresponding
quantum-mechanical systems. The eigenvalues of these quantum systems exhibit
characteristic behaviors that are correlated with those of the associated
classical system.
\end{abstract}
\maketitle

\section{Introduction}
\label{s1}
This paper studies the eigenvalues of the non-Hermitian quantum systems defined
by the $\cP\cT$-symmetric Hamiltonians
\begin{equation}
H=p^2+x^2(ix)^\epsilon\qquad(\epsilon\geq0).
\label{e1}
\end{equation}
The Hamiltonians (\ref{e1}) can have many different spectra depending on the
large-$|x|$ boundary conditions that are imposed on the solutions to the
corresponding time-independent Schr\"odinger eigenvalue equation
\begin{equation}
-\psi''(x)+x^2(ix)^\epsilon\psi(x)=E\psi(x).
\label{e2}
\end{equation}
The boundary conditions on $\psi(x)$ are imposed in Stokes' wedges in the
complex-$x$ plane. At the edges of the Stokes wedges both linearly independent
solutions to (\ref{e2}) are oscillatory as $|x|\to\infty$. However, in the
interior of the wedges one solution decays exponentially and the linearly
independent solution grows exponentially. The eigenvalues $E$ are determined by
requiring that the solution $\psi(x)$ decay exponentially in two nonadjacent
wedges. Ordinarily, the eigenvalues are complex. However, if the two wedges are
$\cP\cT$-symmetric reflections of one another, then there is a possibility that
all the eigenvalues will be real. (The $\cP\cT$ reflection of the complex number
$x$ is the number $-x^*$.)

If $\epsilon$ is not a rational number, there are infinitely many Stokes'
wedges, which we label by the integer $K$. The center of the $K$th Stokes wedge
lies at the angle
\begin{equation}
\theta_{\rm center}=\frac{(4K+2)\pi}{4+\epsilon}-\frac{\pi}{2}.
\label{e3}
\end{equation}
The upper edge of the wedge lies at the angle
\begin{equation}
\theta_{\rm upper}=\frac{(4K+3)\pi}{4+\epsilon}-\frac{\pi}{2}
\label{e4}
\end{equation}
and the lower edge lies at the angle
\begin{equation}
\theta_{\rm lower}=\frac{(4K+1)\pi}{4+\epsilon}-\frac{\pi}{2}.
\label{e5}
\end{equation}
The opening angles of all wedges are the same:
\begin{equation}
{\rm opening~angle}=\frac{2\pi}{4+\epsilon}.
\label{e6}
\end{equation}
The $\cP\cT$ reflection of the wedge $K$ is the wedge $-K-1$. 

\begin{figure*}[t!]
\vspace{5.1in}
\includegraphics{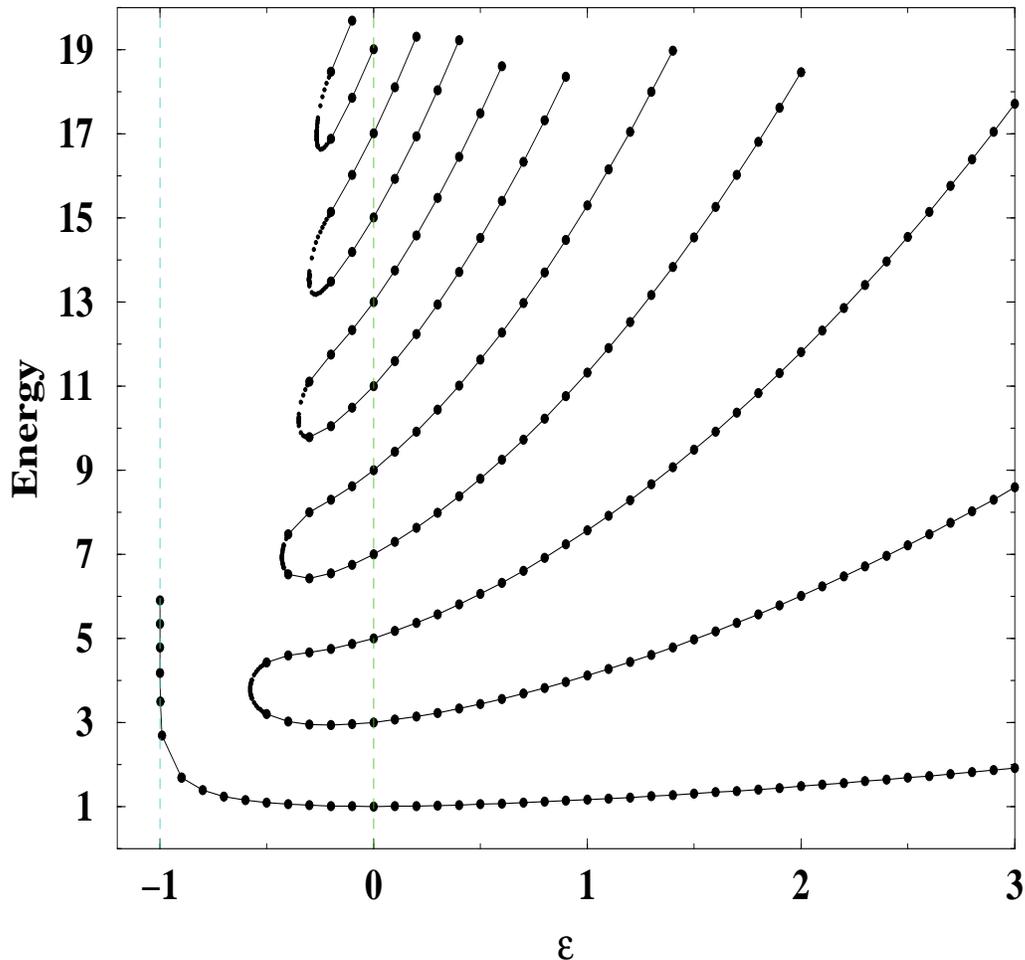}
\caption{Eigenvalues of the Hamiltonian $H$ in (\ref{e1}). These eigenvalues are
found by solving the Schr\"odinger equation (\ref{e2}) with boundary conditions
imposed in the $K=0$ and $K=-1$ wedges. The eigenvalues are real, positive, and
discrete when $\epsilon\geq0$. When $\epsilon=0$ the spectrum consists of the
standard harmonic-oscillator eigenvalues $E_n=2n+1$ ($n=0,\,1,\,2,\,\ldots$). As
$\epsilon$ increases, the eigenvalues grow and become increasingly separated.}
\label{fig1}
\end{figure*}

The $\cP\cT$-symmetric eigenvalue problem posed in the wedges $K=0$ and $K=-1$
has been studied in depth in \cite{R1,R2,R3}. The spectrum for this problem is
real when $\epsilon\geq0$, as illustrated in Fig.~\ref{fig1}. For these boundary
conditions the quantum theory specified by the Hamiltonian exhibits unitary time
evolution \cite{R4}.

The classical system corresponding to this quantum system has also been examined
carefully \cite{R2}. Without loss of generality, one can take the classical
energy to be unity. One can then graph the classical trajectories by solving
numerically the system of Hamilton's differential equations
\begin{equation}
{\dot x}=\frac{\partial H}{\partial p}=2p,\quad
{\dot p}=-\frac{\partial H}{\partial x}=-(2+\epsilon)x(ix)^\epsilon
\label{e7}
\end{equation}
for an initial condition $x(0)$; $p(0)$ is determined from the equation $H=E=1$.
Since $x(0)$ and $p(0)$ are not necessarily real and the differential equations
(\ref{e7}) are complex, the classical trajectories are curves in the complex-$x$
plane. For $\epsilon\geq0$ nearly all trajectories are closed curves. (When
$\epsilon$ is a positive integer, trajectories originating at some turning
points can run off to infinity, but these are special isolated cases. When
$\epsilon<0$, all trajectories are open curves.) If $\epsilon$ is noninteger,
there is a branch cut in the complex-$x$ plane, and to be consistent with $\cP
\cT$ symmetry we take this cut to run from $0$ to $\infty$ along the
positive-imaginary axis. A closed trajectory may cross this branch cut and visit
many sheets of the Riemann surface before returning to its starting point.

As $\epsilon$ increases from 0, the harmonic-oscillator turning points at $x=1$
(and at $x=-1$) rotate downward and clockwise (anticlockwise) into the
complex-$x$ plane. These turning points are solutions to the equation $1+(ix)^{
2+\epsilon}=0$. When $\epsilon>0$ is irrational, this equation has infinitely
many solutions; all solutions lie on the unit circle and have the form
\begin{equation}
x=e^{i\theta_{\rm turning\,point}},
\label{e8}
\end{equation}
where
\begin{equation}
\theta_{\rm turning\,point}=\frac{(2K+1)\pi}{2+\epsilon}-\frac{\pi}{2}.
\label{e9}
\end{equation}
These turning points occur in $\cP\cT$-symmetric pairs (pairs that are symmetric
when reflected through the imaginary axis) corresponding to the $K$ values $(K=-
1,~K=0)$, $(K=-2,~K=1)$, $(K=-3,~K=2)$, $(K=-4,~K=3)$, and so on. When
$\epsilon$ is rational, there are only a finite number of turning points in the
complex-$x$ Riemann surface. For example, when $\epsilon=\frac{12}{5}$, the
Riemann surface consists of five sheets and there are eleven pairs of turning
points.

The period $T$ of a classical orbit depends on the specific pairs of turning
points that are enclosed by the orbit and on the number of times that the orbit
encircles each pair. As explained in Refs.~\cite{R5,R6}, any orbit can be
deformed to a simpler orbit of exactly the same period. This simpler orbit
connects two turning points and oscillates between them rather than encircling
them. For the elementary case of orbits that oscillate between the $K=-1,~K=0$
pair of turning points, the period of the closed orbit is a smoothly decreasing
function of $\epsilon$: 
\begin{eqnarray}
T(\epsilon)=2\sqrt{\pi}\frac{\Gamma\left[(3+\epsilon)/(2+\epsilon)\right]}{
\Gamma\left[(4+\epsilon)/(4+2\epsilon)\right]}\cos\left(\frac{\epsilon\pi}{4+2
\epsilon}\right).
\label{e10}
\end{eqnarray}
To derive (\ref{e10}) we evaluate the contour integral $\oint dx/p$ along a
closed trajectory in the complex-$x$ plane. This trajectory encircles the
square-root branch cut that joins the pair of turning points. We deform the
contour into a pair of rays that run from one turning point to the origin
and then from the origin to the other turning point. The integral along each ray
is a beta function, which is a ratio of gamma functions. Equation (\ref{e10})
holds for all $\epsilon\geq0$.

Complex trajectories that begin at turning points other than the $K=0$ and $K=-
1$ turning points are more interesting. For these trajectories it was shown in
Refs.~\cite{R5} and \cite{R6} that there are three qualitatively distinct
regions of $\epsilon$. For the $K>0$ turning point and the corresponding $-K-1$
turning point, region I extends from $\epsilon=0$ to $\epsilon=1/K$. As
$\epsilon$ increases from 0 in region I, the period of the classical trajectory
is a smooth monotone decreasing function of $\epsilon$ like that in (\ref{e10}).
In region II, where $\epsilon$ ranges from $1/K$ up to $4K$, the period $T(
\epsilon)$ of the classical trajectories is a noisy function of $\epsilon$ and
the classical orbits in this region are acutely sensitive to the value of
$\epsilon$. Small variations in $\epsilon$ can cause huge changes in
the topology and in the periods of the closed orbits. Depending on the value of
$\epsilon$, some orbits have short periods and others have long and possibly
even arbitrarily long periods. In this region there are isolated values of
$\epsilon$ for which the orbits are not symmetric when reflected through the
imaginary axis, and thus these orbits are {\it not} $\cP\cT$ symmetric; such
orbits are said to have spontaneously broken classical $\cP\cT$ symmetry. In
region III, $\epsilon$ ranges from $4K$ up to $\infty$ and the period of the
orbits is again a smooth monotone decreasing function of $\epsilon$.

\begin{figure*}[t!]
\vspace{5.35in}
\includegraphics{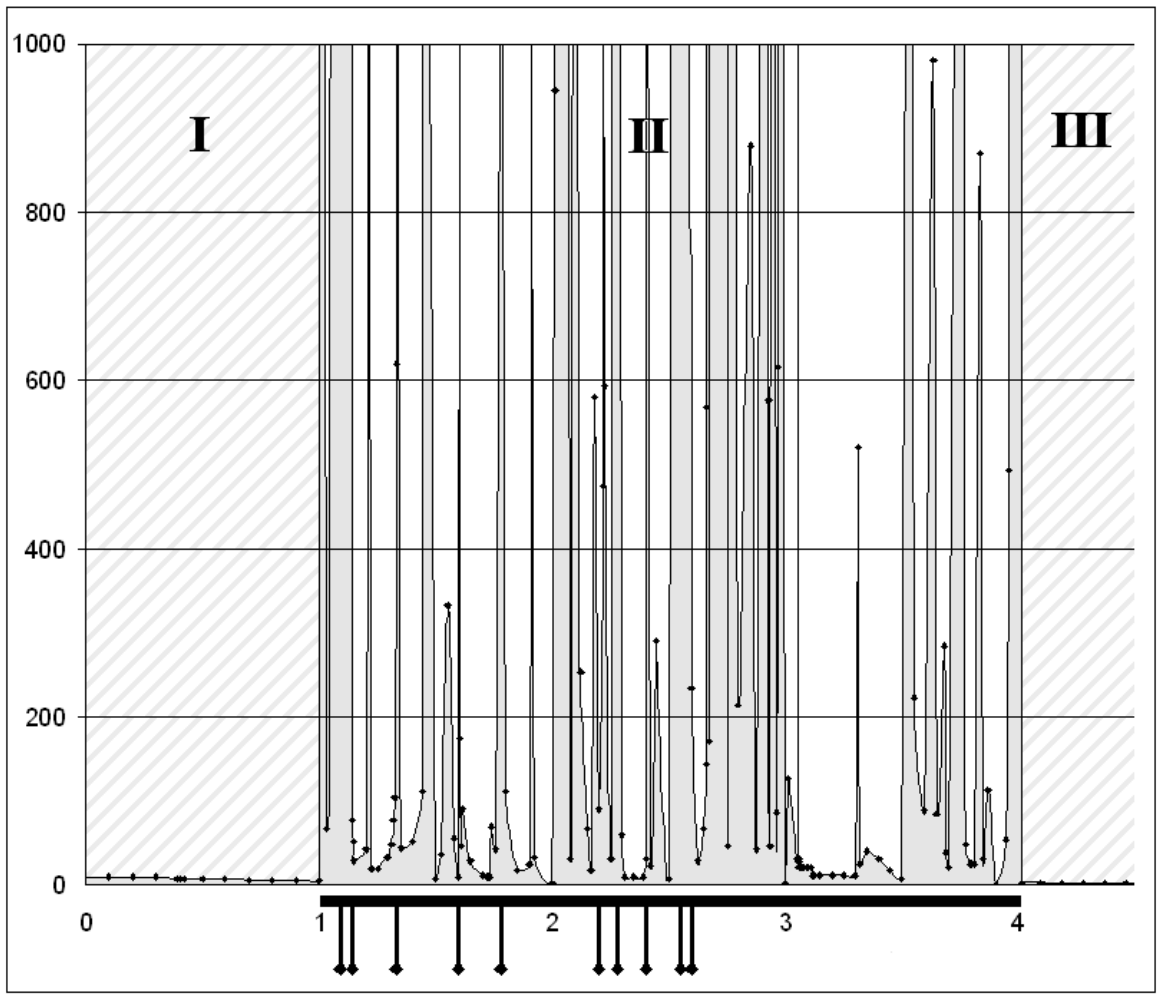}
\caption{Period of a classical trajectory beginning at the $K=1$ turning point
in the complex-$x$ plane. The period is plotted as a function of $\epsilon$. The
period decreases smoothly for $0\leq\epsilon<1$ (region I). However, when $1\leq
\epsilon\leq4$ (region II), the period is a rapidly varying and noisy function
of $\epsilon$. For $\epsilon>4$ (region III) the period is once again a smoothly
decaying function of $\epsilon$. Region II contains short subintervals where the
period is a small and smoothly varying function of $\epsilon$. At the edges of
these subintervals the period abruptly becomes extremely long. Detailed
numerical analysis shows that the edges of the subintervals lie at special
rational values of $\epsilon$ that have the form $p/q$, where $p$ is a multiple
of 4 and $q$ is odd. Some of these rational values of $\epsilon$ are indicated
by vertical line segments that cross the horizontal axis. At these rational
values the classical trajectory does not reach the $K=-2$ turning point and the
$\cP\cT$ symmetry of the classical orbit is spontaneously broken. The periods of
the classical orbits are extremely long near these values of $\epsilon$ in
region II, so in this graph the period in regions I and III appears relatively
small.}
\label{fig2}
\end{figure*}

\begin{figure*}[t!]
\vspace{5.25in}
\includegraphics{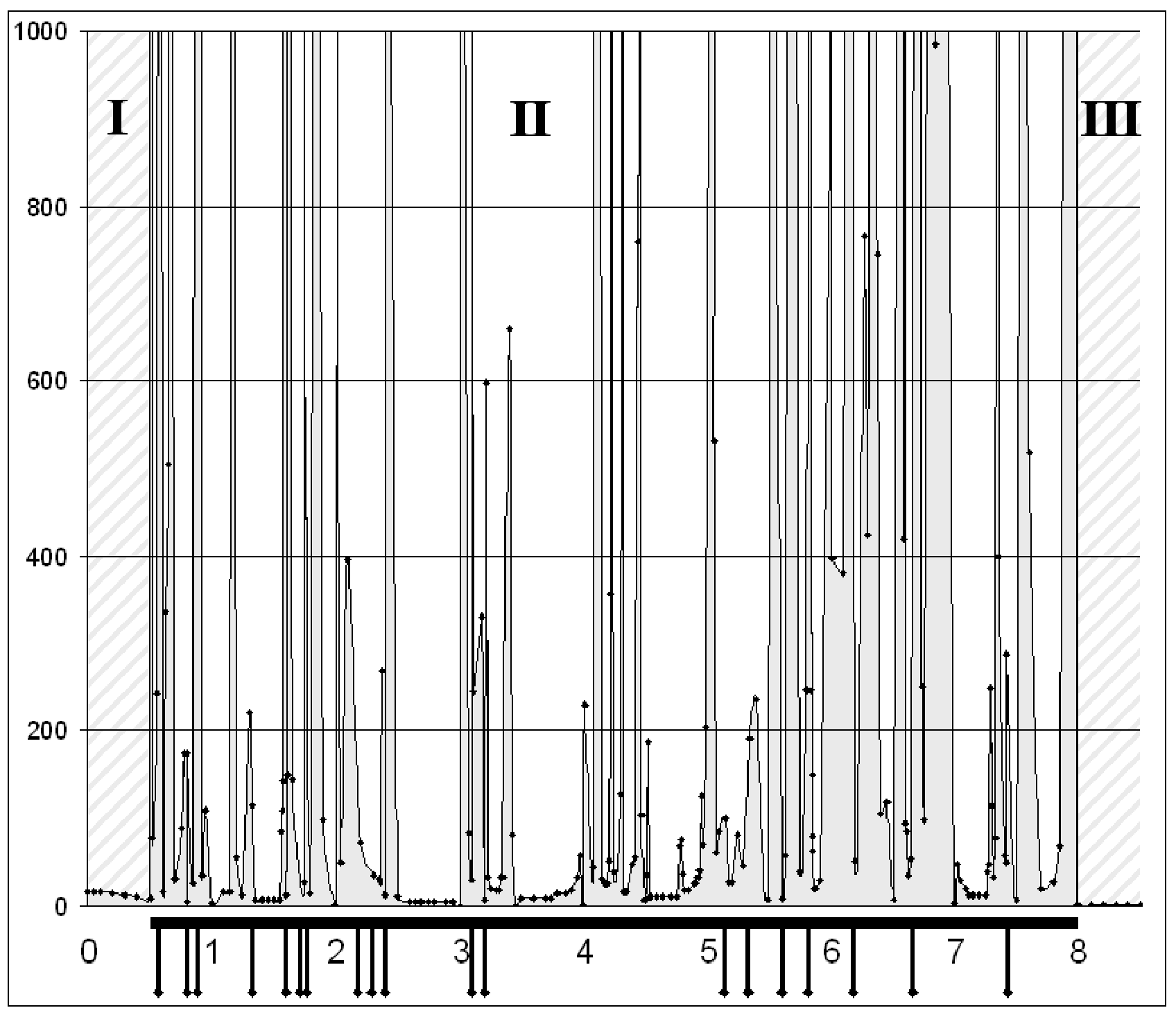}
\caption{Period of a classical trajectory joining (except when $\cP\cT$ symmetry
is broken) the $K=2$ pair of turning points. The period is plotted as a function
of $\epsilon$. As in the $K=1$ case shown in Fig.~\ref{fig2}, there are three
regions. When $0\leq\epsilon\leq\half$ (region I), the period is a smooth
decreasing function of $\epsilon$; when $\half<\epsilon\leq8$ (region II), the
period is a rapidly varying and choppy function of $\epsilon$; when $8<\epsilon$
(region III), the period is again a smooth and decreasing function of
$\epsilon$.}
\label{fig3}
\end{figure*}

The abrupt changes in the topology and the periods of the orbits for $\epsilon$
in region II are associated with the appearance of orbits having spontaneously
broken $\cP\cT$ symmetry. In region II there are short patches where the period
is a relatively small and slowly varying function of $\epsilon$. These patches
are bounded by special values of $\epsilon$ for which the period of the orbit
suddenly becomes extremely long. Numerical studies of the orbits connecting the
$K$th pair of turning points have shown that there are only a finite number of
these special values of $\epsilon$ and that these values of $\epsilon$ are {\it
rational} \cite{R5,R6}. Some special values of $\epsilon$ at which spontaneously
broken $\cP\cT$-symmetric orbits occur are indicated in Figs.~\ref{fig2} and
\ref{fig3} by short vertical lines below the horizontal axis.

Refs.~\cite{R5,R6} explain how very long-period orbits arise. In order for a
classical particle to travel a great distance in the complex plane, its path
must slip through the forest of turning points. When the particle comes under
the influence of a turning point, it usually executes a large number of nested
U-turns and eventually returns back to its starting point. However, for some
values of $\epsilon$ the complex trajectory may evade many turning points before
it eventually encounters a turning point that takes control of the particle and
flings it back to its starting point. Refs.~\cite{R5,R6} conjectured that there
may be special values of $\epsilon$ for which the classical path manages to
avoid and sneak past all turning points. Such a path would have an infinitely
long period. We still do not know if such infinite-period orbits exist.

Ref.~\cite{R5} did not explain why the period of a closed trajectory is such a
wildly fluctuating function of $\epsilon$. However, in Ref.~\cite{R6} it was
shown that for special rational values of $\epsilon$ the trajectory bumps
directly into a turning point that is located at a point that is the complex
conjugate of the point from which the trajectory was launched. This turning
point reflects the trajectory back to its starting point and prevents the
trajectory from being $\cP\cT$ symmetric. Trajectories for values of $\epsilon$
near these special rational values have extremely different topologies and thus
have periods that tend to be relatively long. This explains the noisy plots in
Figs.~\ref{fig2} and \ref{fig3}.

We are not certain whether for each turning point there are a finite or an
infinite number of special rational values of $\epsilon$ for which the classical
orbit has a broken $\cP\cT$ symmetry. The data used to produce Figs.~\ref{fig2}
and \ref{fig3} is far from exhaustive and the bars underneath the horizontal
axes are only the {\it known} examples of broken $\cP\cT$ symmetry. There are
more such special rational values of $\epsilon=\frac{p}{q}$. 

Having reviewed what is known about the classical situation, we come to the
principal subject of this paper, which is to study what happens at the
quantum-mechanical level (that is, how the eigenvalues behave) as $\epsilon$
increases through regions I, II, and III. In Sec.~\ref{s2} we compare the
classical and quantum theories for two cases, $K=1$ (and $K=-2$) and $K=2$ (and
$K=-3$). Our numerical studies show that the eigenvalues exhibit characteristic
changes in behavior that correspond closely with the changes in behavior of the
classical trajectories. In Sec.~\ref{s3} we make some concluding remarks.

\section{Behavior of eigenvalues as $\epsilon$ passes through regions I -- III}
\label{s2}

\subsection{Eigenvalues corresponding with $K=1$ and $K=-2$ turning points}
\label{ss2a}

\begin{figure*}[t!]
\vspace{4.35in}
\includegraphics{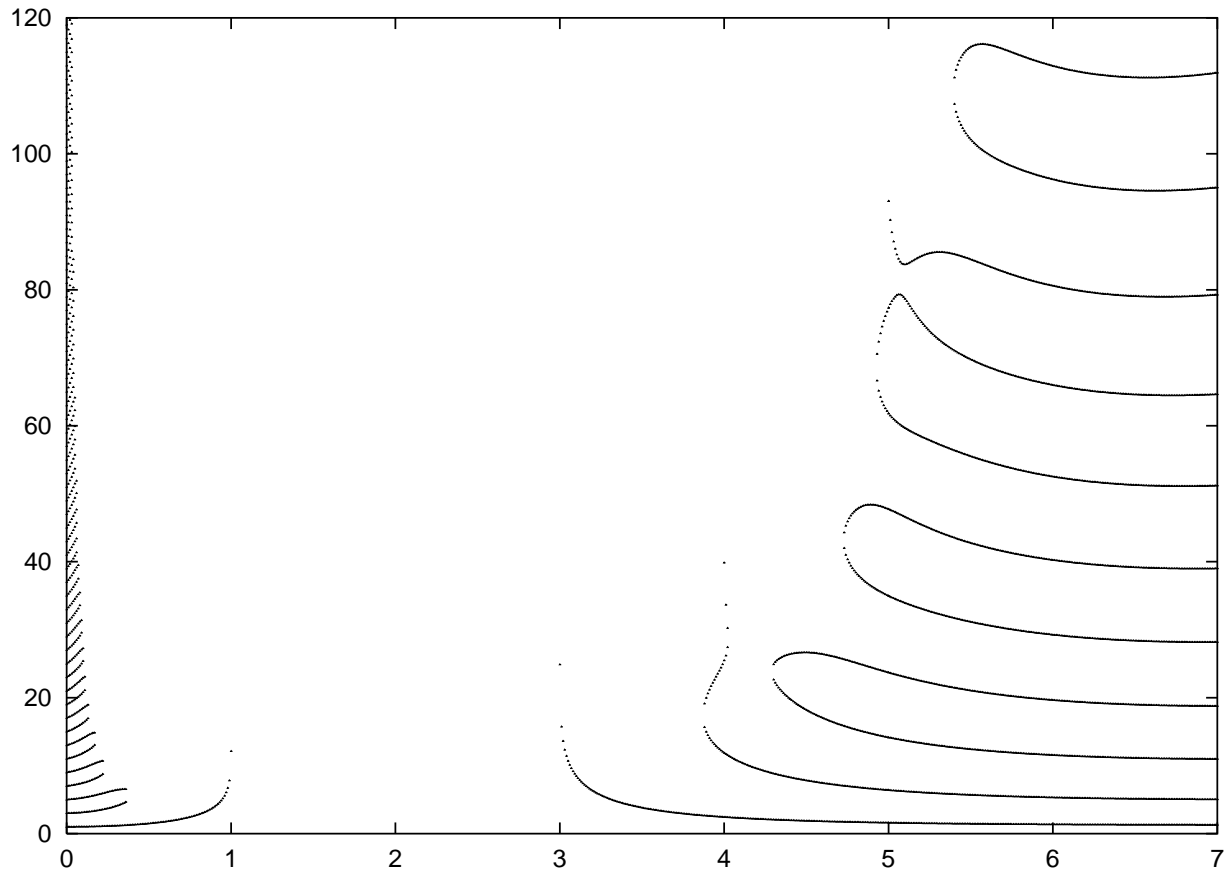}
\caption{Eigenvalues associated with the $K=1$ and $K=-2$ pair of turning points
plotted as functions of $\epsilon$. (The periods of the classical orbits that
begin at the $K=1$ turning point are shown in Fig.~\ref{fig2} for regions I --
III.) When $\epsilon=0$, the spectrum is that of the harmonic oscillator. As
soon as $\epsilon$ increases from $0$, the eigenvalues begin to become complex,
starting with the high-energy part of the spectrum. When $\epsilon=1$, there are
no more real eigenvalues. Real eigenvalues begin to reappear at $\epsilon=3$,
and at the isolated point $\epsilon=4$ all eigenvalues are real. The spectrum is
once again entirely real when $\epsilon$ is greater than about $5.5$.}
\label{fig4}
\end{figure*}

In Refs.~\cite{R5} and \cite{R6} the behavior of the classical trajectories
was studied and the period of the classical trajectories was found to have a
remarkably complicated dependence on $\epsilon$ (see Fig.~\ref{fig2}). In
Fig.~\ref{fig4} we plot the eigenvalues for the corresponding eigenvalue problem
for $0\leq\epsilon\leq7$. (By the corresponding eigenvalue problem, we mean the 
eigenvalues for the eigenvalue problem associated with the $K=1$ and $K=-2$
Stokes wedges.) These eigenvalues were calculated by numerically integrating the
Schr\"odinger equation from large $|x|$ into the origin along the centers of
the two Stokes wedges and then applying the appropriate matching condition on
$\psi(x)$ and $\psi'(x)$ at $x=0$. At the origin $\psi(x)$ is continuous, but
the matching condition on $\psi'(x)$ incorporates the angle of rotation between
the two wedges. These matching conditions are reminiscent of the techniques used
by M.~Znojil in his discussion of ``quantum toboggans" \cite{R7}.

The eigenvalues in Fig.~\ref{fig4} exhibit the following features: At $\epsilon=
0$ the spectrum is that of the harmonic oscillator, $E_n=2n+1$ ($n=0,\,1,\,2,\,
\ldots$), and the entire spectrum is real, positive, and discrete. As soon as
$\epsilon$ increases from 0, the eigenvalues begin to coalesce into complex
conjugate pairs; the process of combining into complex-conjugate pairs starts at
the high-energy part of the spectrum. (This disappearance of real eigenvalues
qualitatively resembles that shown in Fig.~\ref{fig1} as $\epsilon$ decreases
below 0.) When $\epsilon$ reaches the value $0.5$, only one real eigenvalue (the
ground-state eigenvalue) remains. This eigenvalue becomes infinite at $\epsilon=
1$. Between $\epsilon=1$ and $\epsilon=3$ there are no real eigenvalues at all.
The real ground-state eigenvalue reappears when $\epsilon$ increases past $3$,
and as $\epsilon$ approaches $4$, more and more real eigenvalues appear. At the
isolated point $\epsilon=4$ the spectrum is again entirely real and agrees
exactly with the spectrum in Fig.~\ref{fig1} at $\epsilon=4$. At the isolated
point $\epsilon=5$ the spectrum is once again entirely real, but just above and
below this value of $\epsilon$ the spectrum becomes complex again. Finally, as
$\epsilon$ reaches a value near $\epsilon=5.5$, the spectrum is again entirely
real, positive, and discrete and it remains so for all larger values of
$\epsilon$.

Note that some of the qualitative changes in Figs.~\ref{fig2} and \ref{fig4}
occur at the same value of $\epsilon$. For example, the noisy fluctuations in
Fig.~\ref{fig2} begin at $\epsilon=1$, just as the real ground-state eigenvalue
blows up, and the fluctuations stop temporarily at $\epsilon=3$, just as the
real ground-state eigenvalue reappears in Fig.~\ref{fig4}. The noisy classical
fluctuations cease completely at $\epsilon=4$, just as the real eigenvalues
begin to reappear in Fig.~\ref{fig4}.

\begin{figure*}[t!]
\vspace{2.50in}
\includegraphics{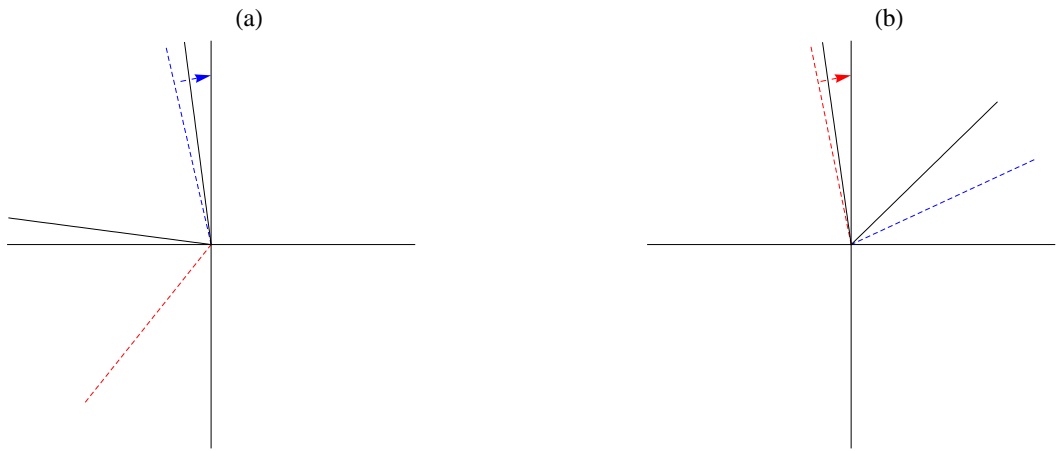}
\caption{Turning points entering and leaving the $K=1$ wedge as $\epsilon$
increases. As $\epsilon$ increases past $1$, the $K=1$ turning point, whose
angle is indicated by a dotted line, leaves the wedge, as shown by the arrow
[panel (a)]. Then, as $\epsilon$ increases past $3$, a new turning point enters
the wedge [panel (b)].}
\label{fig5lr}
\end{figure*}

The most important qualitative changes in Fig.~\ref{fig4} appear to be caused by
the classical turning points entering and leaving the Stokes wedges inside of
which the eigenvalue problem is specified (see Fig.~\ref{fig5lr}). For the case
$K=1$ the upper and lower edges of the wedge are given by
\begin{equation}
\theta_{\rm upper}=\frac{7\pi}{4+\epsilon}-\frac{\pi}{2}
\qquad{\rm and}\qquad
\theta_{\rm lower}=\frac{5\pi}{4+\epsilon}-\frac{\pi}{2}.
\label{e11}
\end{equation}
At $\epsilon=0$ the $K=1$ classical turning point lies in the center of the
wedge. However, as $\epsilon$ increases, the turning point rotates {\it faster}
than the wedge rotates, and at $\epsilon=1$ this turning point leaves the wedge,
as shown in Fig.~\ref{fig5lr}(a). There are no turning points in the wedge
between $\epsilon=1$ and $\epsilon=3$, which corresponds with the total
nonexistence of real eigenvalues. At $\epsilon=3$ a turning point re-enters the
wedge, as shown in Fig.~\ref{fig5lr}(b), but this is the $K=2$ turning point and
not the $K=1$ turning point. The re-entry of a classical turning point is
strongly correlated with the reappearance of real eigenvalues.

When the turning point leaves the wedge, WKB quantization is no longer possible.
Thus, the high-energy portion of the spectrum is no longer real. In fact, we
believe that there are no eigenvalues at all, real or complex, when $1<\epsilon<
3$.

\subsection{Eigenvalues corresponding with $K=2$ and $K=-3$ turning points}
\label{ss2b}

For a classical trajectory beginning at the $K=2$ turning point, the period as a
function of $\epsilon$ again exhibits three types of behaviors. The period
decreases smoothly for $0\leq\epsilon<\half$ (region I). When $\half\leq\epsilon
\leq8$ (region II), the period becomes a rapidly varying and noisy function of
$\epsilon$. When $\epsilon>8$ (region III) the period is once again a smoothly
decaying function of $\epsilon$. These behaviors are shown in Fig.~\ref{fig3}.
As in Fig.~\ref{fig2}, the period in regions I and III is very small compared
with the period in region II. The classical trajectory that begins at the $K=2$
turning point terminates at the $K=-3$ turning point except when $\cP\cT$
symmetry is spontaneously broken. Broken-symmetry orbits occur at isolated
points in region II.

\begin{figure*}[t!]
\vspace{4.45in}
\includegraphics{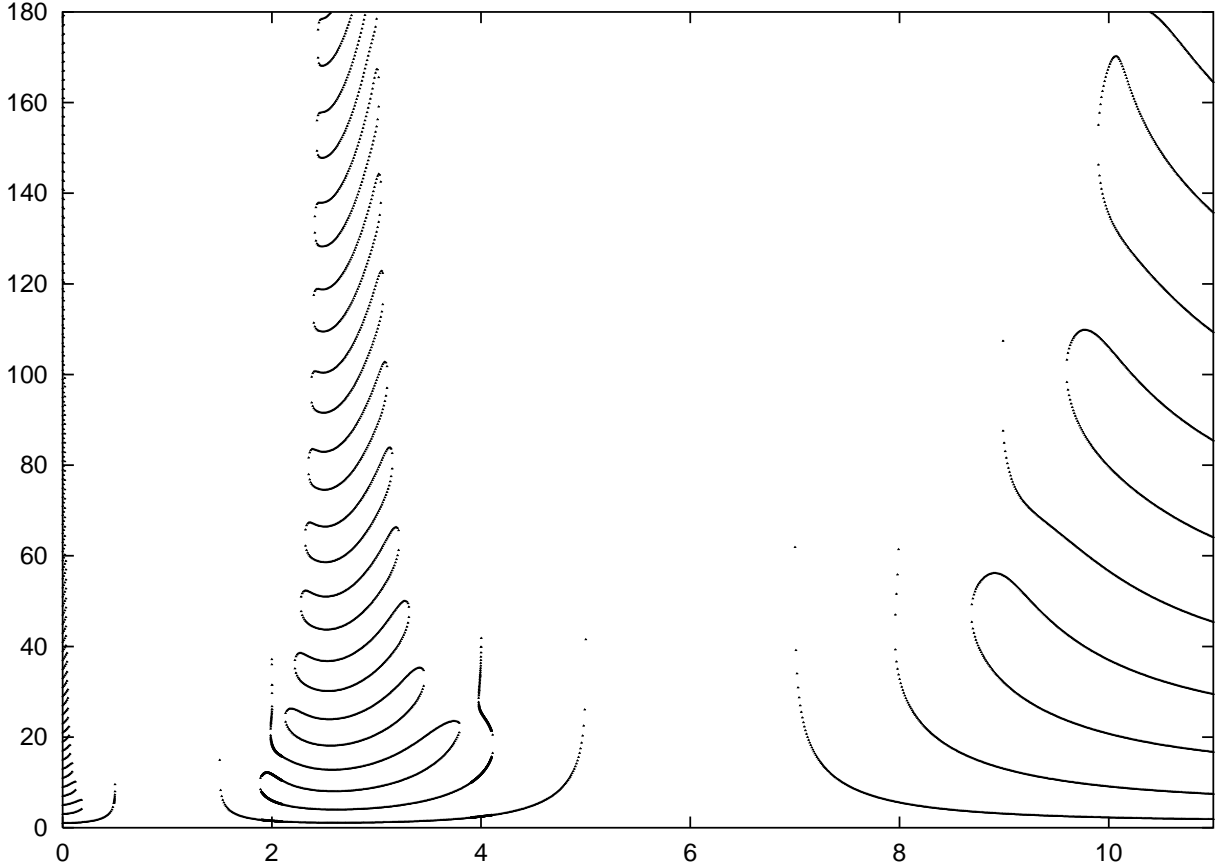}
\caption{The eigenvalues associated with the $K=2$ and $K=-3$ pair of turning
points plotted as a function of $\epsilon$. (The periods of the classical orbits
that begin at the $K=1$ turning point are shown in Fig.~\ref{fig3} for regions I
-- III. When $\epsilon=0$, the spectrum is that for the harmonic oscillator. As 
soon as $\epsilon$ increases from $0$, the eigenvalues begin to become complex, 
starting with the high-energy part of the spectrum, and when $\epsilon=1/2$,
there are no more real eigenvalues. Real eigenvalues begin to reappear at
$\epsilon=3/2$, and there is a region of $\epsilon$ near $5/2$ where the
spectrum is entirely real. There are no real eigenvalues between $\epsilon=5$
and $\epsilon=7$. At the isolated values $\epsilon=2$, $4$, $8$, and $9$ all
eigenvalues are real. The spectrum is once again entirely real when $\epsilon$
is greater than about $10.5$.}
\label{fig6}
\end{figure*}

Figure~\ref{fig6} displays the eigenvalues of the corresponding quantum problem.
Observe that the eigenvalues are all real at $\epsilon=0$ (the harmonic
oscillator spectrum), but they they begin to become complex as $\epsilon$
increases from $0$. There are no real eigenvalues when $\epsilon$ reaches the
value $1/2$, just at the onset of noisy fluctuations in Fig.~\ref{fig3}. There
are several isolated values of $\epsilon$ at $2$, $4$, $5$, $7$, $8$, and $9$,
where the spectrum is entirely real. There are also two regions, $1/2\leq
\epsilon\leq3/2$ and $5\leq\epsilon\leq7$, where there are no real eigenvalues
at all. Between $2$ and $3$ there appears to be a patch where the eigenvalues
are all (or almost all) real. (This patch corresponds with a temporary
disappearance of noise in Fig.~\ref{fig3}). Real eigenvalues begin to reappear
at $\epsilon=9$, and the spectrum appears to be completely real beyond $\epsilon
=21/2$.

The regions in which there are no real eigenvalues are exactly correlated with 
turning points leaving the Stokes wedges. At $\epsilon=1/2$ the $K=2$ classical
turning point leaves the wedge at the angle $\theta=3\pi/2$. Then, at $\epsilon=
3/2$ the $K=3$ classical turning point enters the wedge at the angle $3\pi/2$.
Again, at $\epsilon=5$ the $K=3$ turning point leaves the wedge at the angle
$\pi/2$ and at $\epsilon=7$ the $K=4$ classical turning point enters the wedge
at the angle $\theta=\pi/2$.

\section{Concluding Remarks}
\label{s3}

Figures \ref{fig2} and \ref{fig3} for $K=1$ and Figs.~\ref{fig4} and \ref{fig6}
for $K=2$ illustrate a general pattern that appears to hold for all $K$. For
classical orbits that oscillate between the $K$th pair of turning points, there
are always three regions, region I for which $0\leq\epsilon\leq\frac{1}{K}$,
region II for which $\frac{1}{K}<\epsilon<4K$, and region III for which $4K<
\epsilon$. When $\epsilon=0$, the turning points lie on the real axis in the
centers of the Stokes wedges. As $\epsilon$ increases, the turning points rotate
faster than the wedges rotate and they rotate out of the wedges at $\epsilon=
\frac{1}{K}$. At this point there is a patch where there are no real
eigenvalues. As $\epsilon$ continues to increase, the $K+1$ turning point
rotates into the Stokes wedge and the eigenvalues begin to become real.
Eventually, this turning point leaves the wedge and there are again no real
eigenvalues. In total, there are $K$ regions where there are no real
eigenvalues. Eventually, the $2K$ turning point enters the wedge at the end of
region II, and the spectrum begins to become entirely real and remains entirely
real in region III.

\begin{figure*}[t!]
\vspace{2.70in}
\includegraphics{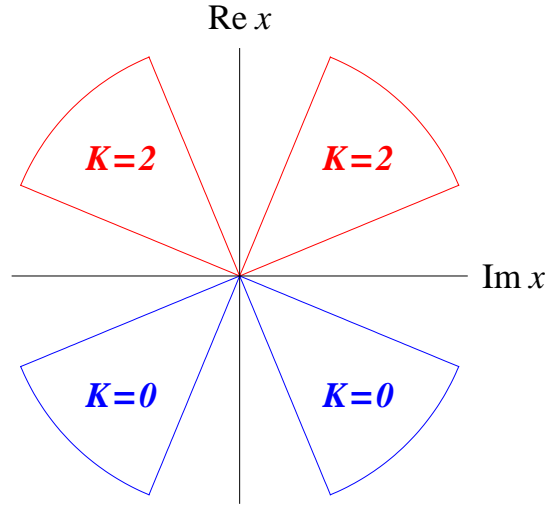}
\caption{The Stokes wedges for $\epsilon=4$ for the cases $K=0$ and $K=2$.
Because the wedges are complex conjugate pairs the spectra are identical for
this value of $\epsilon$.}
\label{fig7}
\end{figure*}

There are many isolated points at which the eigenvalues are all real. For
example, when $K=2$, real eigenvalues occur at $2$, $4$, $8$, and $9$. Some of
these points can be analyzed in terms of $\cP\cT$ reflection. Specifically:
(i) From (\ref{e3}) we can see that at $\epsilon=2$ the $K=2$ wedge is centered
at the angle $7\pi/6$. The $\cP\cT$ reflection of this wedge lies at the angle
$-\pi/6$. Thus, the $K=2$ spectrum agrees exactly with the $K=0$ spectrum in
Fig.~\ref{fig1} for this value of $\epsilon$. (ii) At $\epsilon=4$ the $K=2$
wedge is centered at $3\pi/4$. The $\cP\cT$ reflection of this wedge is
centered at $\pi/4$, and under complex conjugation this angle becomes $-\pi/4$.
Thus, the $K=2$ spectrum is identical to the $K=0$ spectrum for this value of
$\epsilon$. This configuration of wedges is shown in Fig.~\ref{fig7}. (iii) At
$\epsilon=8$ the $K=2$ wedge is centered at $\pi/3$. Under complex conjugation,
this wedge is now centered at the angle $-\pi/3$, which corresponds with the
wedge for the $K=0$ case. Thus, once again, the spectra for these two problems
are identical at this value of $\epsilon$. (Note that for these three
situations $\epsilon$ is integer valued, and hence there is no cut in the
complex-$x$ plane.) 

While quantum mechanics and classical mechanics are of course different
theories with different regimes of applicability, our investigation has shown
that there are nonetheless some correlations between the two. Regular classical
behavior seems to correspond to a completely real quantum spectrum, while
chaotic classical behavior in the complex plane is associated with either a
partially complex spectrum or a complete absence of eigenvalues. In the systems
we have studied, the onset of rapid and irregular variation in the periods
of the classical trajectories correlates precisely with the disappearance of all
quantum eigenvalues as the ground-state energy goes to infinity, while the end
of the noisy region is associated with the gradual reappearance of a completely
real spectrum. Within the region of rapid variation some correlations can be
observed, but they are less clear-cut and merit further analysis.

\begin{acknowledgments}
We thank D.~Darg for helpful discussions and D.~Hook for programming assistance.
CMB is supported by a grant from the U.S.~Department of Energy.
\end{acknowledgments}


\begin{thebibliography}{99}

\bibitem{R1} C.~M.~Bender and S.~Boettcher, Phys.~Rev.~Lett.~{\bf 80}, 5243
(1998).

\bibitem{R2} C.~M.~Bender, S.~Boettcher, and P.~N.~Meisinger,
J.~Math.~Phys.~{\bf 40}, 2201 (1999).

\bibitem{R3} P.~Dorey, C.~Dunning, and R.~Tateo, J.~Phys.~A: Math.~Gen.~{\bf
34}, L391 (2001) and {\bf 34}, 5679 (2001).

\bibitem{R4} C.~M.~Bender, D.~C.~Brody, and H.~F.~Jones, Phys.~Rev.~Lett.~{\bf
89}, 270401 (2002).

\bibitem{R5} C.~M.~Bender, J.-H.~Chen, D.~W.~Darg, and K.~A.~Milton,
J.~Phys.~A: Math.~Gen.~{\bf 39}, 4219 (2006).

\bibitem{R6} C.~M.~Bender and D.~W.~Darg, J.~Math.~Phys.~{\bf 48}, 042703
(2007).

\bibitem{R7} M.~Znojil, J.~Phys.: Conference Series {\bf 128}, 012046 (2008).

\end{thebibliography}
\end{document}